\documentclass[pra,aps,twocolumn,eqsecnum,showpacs]{revtex4}

\usepackage{graphics}

\begin{document}

\title{Spectral dependence of diffuse light dynamics in ultracold atomic $^{85}$Rb}

\author{S. Balik, R. Olave, C.I. Sukenik, and M.D. Havey
\break Department of Physics, Old Dominion University, Norfolk, VA
23529 \break \break V.M. Datsuk, D.V. Kupriyanov and I.M. Sokolov
\break Department of Theoretical Physics, State Polytechnic
University, 195251, St.-Petersburg, Russia}

\begin{abstract}
We report a combined experimental and theoretical simulation of
multiply scattered light dynamics in an ultracold gas of $^{85}$Rb
atoms. Measurements of the spectral dependence of the time-decay
of the scattered light intensity, following pulsed excitation with
near resonance radiation, reveals that the decay for long times is
nearly exponential, with a decay constant that is largely
independent of detuning from resonance. Monte Carlo simulations of
the multiple scattering process show that, for large detunings,
near resonance scattering of Fourier components of the excitation
pulse plays a significant role in the effect. This interpretation
is supported by the observations, and successful modelling, of
beating between Rayleigh scattered light at the excitation carrier
frequency with the Fourier components of the excitation pulse that
overlap significantly with the atomic resonance.
\end{abstract}

\maketitle

\section{Introduction}
Multiple scattering of near-resonance radiation is well known to
have significant effects on the density distribution of trapped
ultracold atoms \cite{Metcalf}, and on the collective dynamics of
the trapped ensemble itself \cite{LabeyrieOscillations}. However,
in spite of the enormous number of applications and studies of
atoms confined in a MOT, there have been relatively few direct
studies of the dynamics of the multiply scattered light itself.
The limited number of studies include measurements of radiation
trapping \cite{Holstein} of resonance radiation in an ultracold
gas of atomic Cs by Fioretti, \emph{et al.} \cite{Fioretti} and
studies of the long-time dynamics of light scattered in the
vicinity of the atomic $^{85}$Rb $F = 3 \rightarrow F' = 4$
resonance transition \cite{LabeyrieTime1,LabeyrieTime2}; these
measurements were also made in an ultracold gas, for which there
is limited frequency redistribution of the scattered light. Recent
measurements by Balik, \emph{et al.} \cite{Balik1} examined the
influence of quantum interference on the dynamics of the atomic
alignment produced in the vicinity of the $F = 3 \rightarrow F' =
4$ hyperfine transition associated with the atomic $^{85}$Rb D2
resonance transition.

Multiple light scattering is also important in several specific
areas of current research interest
\cite{Sheng,LagTig,Wiersma1,Chabanov1,Mish}. These areas are
focussed on the role of interferences in multiple wave scattering
in a wide variety of condensed, liquid, or gaseous samples.  Such
interferences are readily observable in many condensed systems and
under a broad range of circumstances.  For warm atomic gases, the
interferences are normally not observable because of frequency
redistribution caused by the thermal motion of the constituent
atoms or molecules.  However, in ultracold atomic gases
\cite{Labeyrie1,Labeyrie2,Antilocalization,HaveyCBS}, where the
thermal motion of the atoms is very small, the interferences
become readily observable through the coherent backscattering
effect \cite{Ishimaru,Wolf,Albeda}. Research to date has recently
been reviewed by Havey, \emph{et al.} \cite{CommentAMO} and by
Kupriyanov, \emph{et al.} \cite{LPRReview}. In relatively low
density gases, where the mean free path for light scattering is
much greater than the mean separation between the atoms, the
process can be considered to be due to a sequence of separate
scattering and propagation events. All studies to date have been
done in the so-called weak localization regime, defined by the
Ioffe-Regel condition as occurring when k\emph{l} $>>$ 1. Here $k$
is the wave vector of the light, while \emph{l} is the mean free
path for light propagation in the medium.  The coherent
backscattering enhancement is due to distributions of chains of
such sequences with a corresponding distribution of scattering
orders. Scattering along reciprocal paths then leads to the
interferometric enhancement of light scattered in the
backscattering direction; in this direction the relative
geometrical phase of the reciprocal paths is preserved,
irrespective of the location of the atoms comprising the chains.
When the density of atoms is increased to the point where the mean
free path for light scattering is \emph{on the order of} the
atomic separation, then recurrent scattering \cite{Sheng} becomes
important, and a transition from propagating light modes to
localized ones becomes possible.  The boundary near which this is
expected to become important is defined by the Ioffe-Regel
condition as k\emph{l} $\sim$ 1.  Localization by disorder (in
this case disorder due to the random distribution of atoms in the
sample) is the optical analog of Anderson localization
\cite{Anderson} of electrons.  One driving motivation for studies
of diffusive light propagation is to understand the purely
diffusive part of the light propagation, and how the contrast with
a transition to localized modes might be observed.  Light
localization has been reported for condensed samples in the
optical \cite{Wiersma1} and in the microwave regimes
\cite{Chabanov1}, but not in an atomic gas.

A second area of considerable interest is coherent manipulation of
light propagation in ultracold atomic gases \cite{Lukin1}.  This
is accomplished by using effects associated broadly with coherent
population trapping \cite{Arimondo1,Gray,Yoo}, and specifically
under conditions associated with electromagnetically induced
transparency (EIT) \cite{Arimondo2, Harris, Marangos, Wynands}.
One remarkable demonstration of this was the seminal experiment by
Hau, \emph{et al.} \cite{Slowlight}, in which reduction of the
speed of a coherent beam of resonance radiation to 17 m/s in
ultracold atomic sodium was observed. In this experiment, the
group velocity of a weak probe beam was manipulated by a second
control beam, where the two light sources were tuned to a
two-photon Raman resonance between Na ground state hyperfine
levels.  Other experiments have used EIT to demonstrate nonlinear
optical effects including four-wave mixing \cite{Braje}, optical
information storage \cite{Liu2}, remarkable recent experiments
demonstrating generation and control of single photons
\cite{Braje2, Chaneliere, Eisaman}, and general studies of EIT
phenomenology in ladder and lambda type configurations \cite{Yan}.
All such studies have been concerned with single or multiple
coherent beams either applied or generated by nonlinear processes
in the samples. However, as demonstrated by the coherent
backscattering effect, phase can also be preserved in multiple
scattering of light in an ultracold atomic gas \cite{note}. This
leads to the possibility that the CBS effect, in the weak
localization regime, can be modified by the presence of a second
control field in one of the customary EIT configurations. Recent
theoretical results have in fact demonstrated that profound
effects on the CBS intensity profile, spectral variation, and the
temporal pulse shape of multiply scattered light beams are greatly
modified by the presence of a control beam \cite{Sokolov1}. A
physical way to envision these effects is to realize that the CBS
effect reflects the optical transport properties of a weak probe
beam, and the presence of an additional control field modifies
these transport properties. This modification of the diffuse light
propagation in an EIT configuration is a newly discovered effect,
and much exploration of the experimental and theoretical landscape
remains to be done.

\section{Experimental Approach} The experimental details of our approach
have been described elsewhere \cite{Balik1}, and will be only
briefly sketched here. The basic experimental scheme is portrayed
in Fig. 1, where it is shown that physical sample is an ultracold
gas of atomic $^{85}$Rb prepared in a magneto optical trap. The
trap operates on the $F = 3 \rightarrow F' = 4$ hyperfine
transition. The trap \cite{HaveyCBS} produces a nearly Gaussian
cloud of $\sim$ $10^{8}$ ultracold rubidium atoms at a temperature
$\sim$ 100 $\mu$K. The peak density is $\sim$ $3$ x $10^{10}$
$cm^{-3}$. Fluorescence imaging is used to estimate the Gaussian
radius of the sample to be $r_{0}$ $\sim$ $1 mm$. The peak optical
depth, as measured by absorption measurements of light transmitted
through the central portion of the atomic sample, is $b_{0}$ =
8(1). For a Gaussian atom distribution in the trap, the weak-field
optical depth, on resonance and through the center of the trap, is
given by $b_{0}$ = $\sqrt{2\pi}$$n_0$$\sigma_{0}$$r_{0}$, where
$n_0$ is the peak trap density and $\sigma_{0}$ is the
on-resonance cross-section. The single resonance scattering cross
section $\sigma$ varies with probe frequency, $b = b_{0}[1 +
(2\Delta/\gamma)^{2}]^{-1}$, where $\Delta = \omega_{L} -
\omega_{0}$, and $\omega_{L}$ is the probe frequency, $\omega_{0}$
is the $F = 3 \rightarrow F' = 4$ hyperfine transition frequency,
and $\gamma$ is the natural width. A continuous wave, low
intensity diode laser serves as probe of the optical transmission
of the MOT and also as the excitation source for the time resolved
experiments reported here.  The laser, which has a bandwidth
$\sim$ 1 MHz, is tuned in a range of several $\gamma$ around this
transition. The average light intensity for the probe is $\sim$ 1
$\mu W/cm^{2}$. To produce a nearly Gaussian beam profile, the
laser output is passed through a single-mode optical fiber and
expanded to a $1/e^{2}$ width $\sim$ 8 mm. The probe laser
intensity is modulated with an acousto optic modulator (AOM),
which generates sharp rectangular pulses having an \emph{on} time
of 2 $\mu s$ and an \emph{off} time of 2 $ms$. The 2 $\mu s$
excitation pulse is centered in a 90 $\mu s$ window during which
fluorescence data are recorded. The MOT lasers are off during the
data acquisition period. For the remaining nearly 2 $ms$, the MOT
lasers are turned back on to regenerate the ultracold atomic
sample. An important component of the experiment is a synchronized
mechanical chopper, which prevents intense fluorescence from the
MOT region, present during the MOT build up period, from reaching
the photomultiplier tube (PMT). The AOM-limited 20 dB response is
estimated to be less than $\sim$ 60 ns. The probe laser is
linearly polarized in the vertical direction.

\begin{figure}
\includegraphics{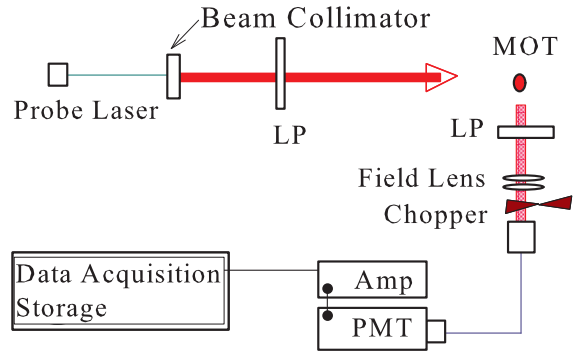}
\caption{Schematic diagram of the experimental arrangement. Shown
in the figure is the probe laser, magneto optic trap (MOT), linear
polarizers (LP), and a photomultiplier tube (PMT) used for
detection of the fluorescence signals.} \label{Figure 1}
\end{figure}

Fluorescence signals are detected in a direction orthogonal to the
probe laser propagation and polarization directions.  To minimize
reduction in the measured light polarization, the scattered light
is collected in a small effective solid angle of about 0.35 mrad,
and refocussed to match the numerical aperture of a 400 $\mu$m
multimode fiber. A linear polarization analyzer is used to select
the detected polarization channel, which we label as parallel
($\|$) and perpendicular ($\bot$). The fiber output is filtered by
a 5 nm spectral width interference filter centered at 780 nm, and
passed to a GaAs-cathode photomultiplier tube (PMT). The PMT is
operated in a photon counting mode, whereby the output is
amplified and directed to a discriminator and multichannel scalar
(MCS) (indicated in Fig. 1 as data acquisition and storage). The
multichannel scaler time sorts and accumulates the data in 5 ns
bins. Timing of the experiment, including the MOT and probe
lasers, and triggering the start of the MCS sweep is controlled by
a precision pulse generator slaved to a timing signal from the
mechanical chopper.

\section{Theoretical Approach}
We provide here only an outline of our theoretical approach to
description of the time evolution of light scattering in an
ultracold atomic gas.  Details of this treatment may be found in
our earlier papers
\cite{Balik1,HaveyCBS,CommentAMO,LPRReview,Sokolov1} The time
dependence of the scattered pulse intensity can be extracted by
using the following calculation steps. Basically the process is
fully described by the scattering amplitude defined for each
spectral component of the incoming and outgoing pulse. Thus, this
scattering amplitude has to be established as a first step. Such
an amplitude for the multiple scattering process developing in an
opaque medium includes two important calculation ingredients,
namely, the scattering tensor and the mesoscopic Green function.
The description of these theoretical characteristics of the
scattering problem for the relevant case of an ultracold atomic
sample can be found in Ref.[\cite{LPRReview}]. Roughly speaking
the amplitude can be constructed for any randomly chosen chain of
atomic scatterers as a subsequent product of the respective
scattering tensors responsible for the deflection of the light ray
from its coherent forward propagation. The transformation of the
field amplitude inside such a zigzag-path, but along the fragments
of freely propagation, can be described in terms of the Green
function formalism.

In the next step the partial contribution to the time profile of
the outgoing light pulse is recovered via a reverse Fourier
transform. This transform should be subsequently evaluated for
each order of the multiple scattering. This part of the simulation
procedure is the most difficult and needs an accumulation of the
quite extended database for the large number of the contributing
spectral components, see \cite{Sokolov1} for details. Finally the
partial contribution to the instantaneous intensity of the light
scattered by the selected chain is expressed by the squared time
dependent partial amplitude. The crucial point of the final
calculational step comes from the important physical requirement
to the interference contribution associated with the scattering in
backward direction. For this particular scattering channel the
partial amplitudes responsible for the light propagation along two
reciprocal scattering path can interfere and the interference
survives the configuration averaging. At the final step the normal
"ladder"-type and, in the case of the backscattering channel, the
"crossed"-type amplitude products should be averaged with the
Monte Carlo technique for a Gaussian-type distribution of the
atomic scatterers in the sample. The total outgoing pulse is
finally expressed by the sum of such partial contributions, and
the sum is typically converged for the orders of multiple
scattering $\sim b_0$.

The described procedure of numerical simulation allows us to
include in the numerical routine most of the practically important
factors such that the hyperfine and Zeeman structure of the ground
and excited states, the spatial inhomogeneity of atomic cloud and
its shape variation. In addition, the dynamical modification of
the scattering process caused by atomic motion can be traced via
the time and spectral dependencies of the correlation function of
the scattered light. Some interesting and important effects for
the light scattering on spin-oriented atomic systems were
predicted with such a detail numerical analysis, see
\cite{LPRReview}.

\section{Results and Discussion} In an earlier paper we reported
on the time dependence of the linear polarization degree of the
multiply scattered light.  In the measurements, the ultracold
atoms are excited by a nearly rectangular 2 $\mu$s long pulse of
linearly polarized resonance radiation tuned in the spectral
vicinity of the $F = 3 \rightarrow F' = 4 \rightarrow F = 3$
transition. The measured intensities of linearly polarized
scattered light in two orthogonal directions (along and
perpendicular to the direction of the exciting light pulse
polarization) may be quantified by defining a linear polarization
degree as

\begin{equation}
P_{L} = \frac{I_{\parallel} - I_{\perp}}{I_{\parallel} +
I_{\perp}}
\label{PL}%
\end{equation}

In the formula, $I_{\parallel}$ and $I_{\perp}$ represent the
measured intensities in the lin $\parallel$ lin and lin $\perp$
lin channels.  It was shown by Balik, \emph{et al.} \cite{Balik1}
that the linear polarization degree decays rapidly towards zero on
a time scale of less than 100 ns. In the present paper, we are
concerned with the longer time dynamics, for which $I_{\parallel}$
and $I_{\perp}$ very nearly equal.  Our results are then presented
solely in terms of $I_{\parallel}$.

\begin{figure}
\includegraphics{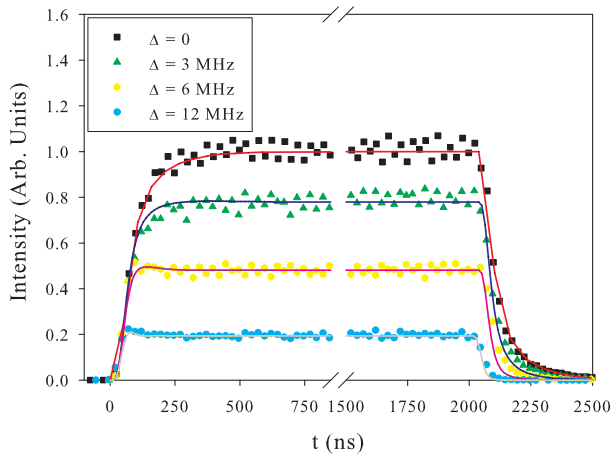}
\caption{Time-dependence of the scattered light intensity, in the
$I_{\parallel}$ polarization channel, for several different
detunings from atomic resonance.} \label{Figure 2}
\end{figure}

The time dependence of the scattered light intensity
$I_{\parallel}$, for several different detunings from the $F = 3
\rightarrow F' = 4$ transition is shown in Fig. 2.  There we see
that, after a transient growth period of approximately 100 ns, the
fluorescence intensity reaches a steady state value.  Upon shut
off of the excitation pulse at a time t = 2000 ns, the
fluorescence decays to a small level on a time scale of several
hundred ns. As the frequency of excitation is shifted away from
the resonance transition, both the transient build up and decay
become more rapid.  This qualitatively makes sense, for the
optical depth of the vapor is reduced with detuning from
resonance, and the radiation thus undergoes fewer scattering
events before emerging from the optically dense sample.  As
indicated by the solid lines in Fig. 2, the overall time
dependence is well described by the results of Monte Carlo
simulations.

The behavior of the intensity after the excitation pulse is turned
off is seen more clearly in the semilogarithmic plots shown in
Figure 3.  There we see a surprising result:  the long time decay
is nearly exponential and has essentially the same decay rate,
regardless of the detuning.  We point out that, even though this
data is for positive detunings of $\Delta$ = 0, 3, 6, and 12 MHz,
a very similar effect is seen for negative detunings of the same
magnitude.  It is also evident that the Monte Carlo simulations,
shown in this graph as solid lines, do not account for the effect
just described.  This suggests that additional physics must be
included to better describe the physical circumstances of the
experiment.

\begin{figure}
\includegraphics{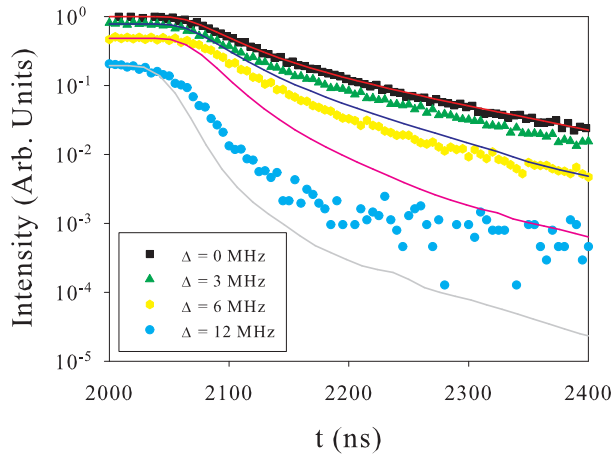}
\caption{Semilogarithmic graph of the time-dependence of the
scattered light intensity, in the $I_{\parallel}$ polarization
channel, for several different detunings from atomic resonance.}
\label{Figure 3}
\end{figure}

One possibility is that the finite bandwidth of about 1 MHz of the
excitation laser results in resonance excitation with the wings of
the laser spectral profile.  Simulations of this effect for laser
widths of 0.01 MHz, 1.0 MHz, and 3.0 MHz are shown in Fig. 4.
There it is seen that for small detunings ($\Delta$ = 3 MHz in
this case), there is some influence, and the change goes in the
proper direction in comparison with the experimental data.
However, the bandwidth effect becomes increasingly negligible for
larger detunings, and so cannot account for the results of Fig. 3.

\begin{figure}
\includegraphics{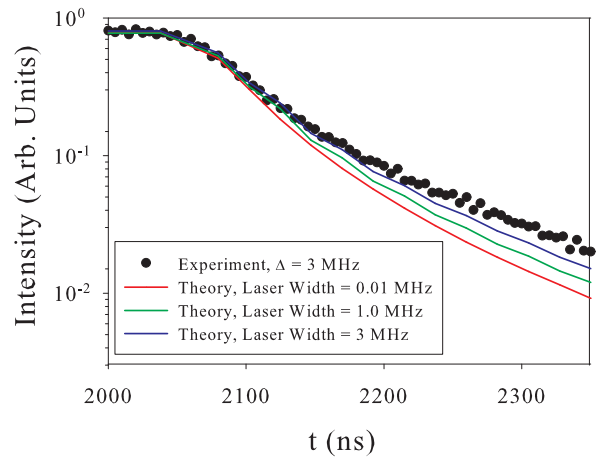}
\caption{Influence of the laser linewidth on the scattered light
intensity, in the $I_{\parallel}$ polarization channel, for a
detuning $\Delta$ = + 3 MHz.} \label{Figure 4}
\end{figure}

The most promising possibility is that spectral components of the
excitation pulse, generated by the relatively rapid turn on and
turn off of the pulse, sufficiently overlap the atomic resonance
transition to explain the effect.  In Fig. 5 we compare the
experimental data at a larger detuning of $\Delta$ =  + 12 MHz
with simulations for pulses having different turn-on and turn-off
models and time constants.  There it is seen, for example, that
for an exponential turn off shape with a time constant on the
order of a few ns, the data and simulations are in reasonable
accord. As the Fourier components for a rectangular pulse drop off
with frequency relatively slowly with offset from the carrier
frequency, this is a suggestive interpretation of the
observations.  We point out that this interpretation is supported
by the transient turn on behavior, as shown in Figure 6.  There we
see a sharp overshoot, which we interpret as beating between the
Rayleigh scattered light and the Fourier components of the
excitation pulse scattered according to the Lorentzian response of
the atomic resonance.  Although the experimental data is too noisy
to extract the beat frequency, the simulations, on an expanded
intensity scale, show that the beat note frequency is in fact
given by the detuning from resonance.

\begin{figure}
\includegraphics{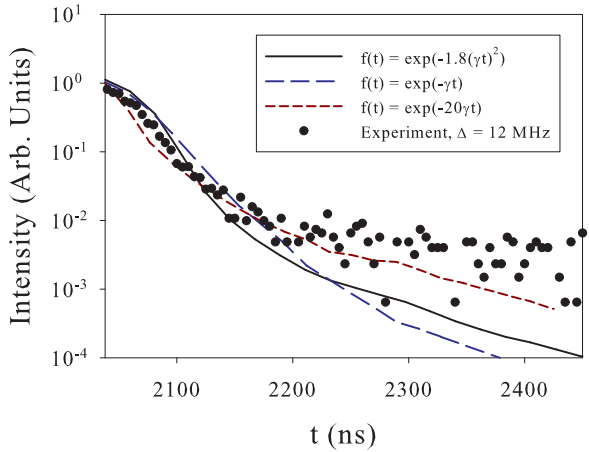}
\caption{Influence of the
temporal shape of the excitation pulse on the long time behavior
of the scattered light intensity in the $I_{\parallel}$
polarization channel. $\Delta$ = + 12 MHz.} \label{Figure 5}
\end{figure}

\begin{figure}
\includegraphics{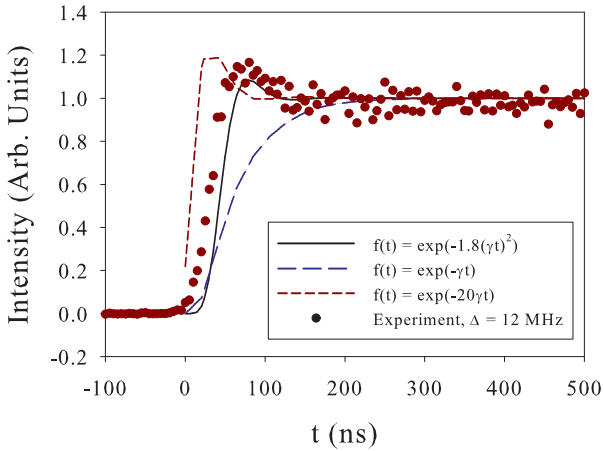}
\caption{Influence of the temporal shape of the excitation pulse
on the short-time growth of the scattered light intensity in the
$I_{\parallel}$ polarization channel. $\Delta$ = + 12 MHz.}
\label{Figure 6}
\end{figure}

Finally, we point out that Labeyrie, \emph{et al.}
\cite{LabeyrieTime1,LabeyrieTime2} have observed a similar effect
in $^{85}$Rb but in a somewhat different fluorescence geometry,
sample geometry, and preparation. In these studies, the Gaussian
radius of the MOT is 2 - 3 mm, while the peak optical depth
$b_{0}$ ranges from 2 to 40.  In addition, optical excitation of
the sample was done with a probe laser smaller than the sample
size, in contrast to the experiments reported here.  Finally, the
angle of observation of the fluorescence was about 17$^{o}$ from
the forward direction, in comparison to our measurements where
this angle is $\sim$ 90$^{0}$.  They interpret the long time
constant as being due to redistribution of the multiply scattered
light by the residual thermal motion of the atoms in the ultracold
atomic cloud. Reported simulations are also in good agreement with
their temperature dependent measurements of the decay rate at
longer times. No data on the turn on portion of the fluorescence
measurements were reported.  In comparing the results of Labeyrie,
\emph{et al.} \cite{LabeyrieTime1,LabeyrieTime2} we see that the
essential effect, that the longer time decay is nearly constant
and independent of detuning, is the same in both experiments.
However, the interpretation of the results is different. This
could well be due to the quite different experimental arrangements
and conditions for the two experiments, under which different
physical mechanisms may be important.  For the experiments
reported here, it seems that the interpretation based on
overlapping Fourier components provide a self consistant
explanation for both the longer time decay and for the shorter
time turn on transient. However, we should point out that
initially warmer atoms also have a larger overlap with the Fourier
components of the excitation pulse, and can contribute to the long
time tail. This should have a prompt effect on the transient
buildup and decay, rather than an time-accumulating effect such as
frequency redistribution, due to thermal motion, in multiple
scattering.  Finally, the data reported in Fig. 5, for example,
show not ideal agreement with our simulations, even accounting for
the Fourier components of the excitation pulse.  This residual
difference may be due to thermal redistribution, which has not
been taken into account in our simulations.

\section{Conclusions}
In conclusion, we report experimental and theoretical results on
the long time decay of near resonance radiation from an ultracold
gas of Rb atoms.  The experimental results show two features
suggesting that Fourier components of the excitation pulse play an
important role in the observed constancy of the scattered light
intensity at longer times following pulse shut off.  Our
simulations also show that realistic laser spectral widths can
influence the observations for smaller detunings from resonance.
Measurements simulations by other researchers, on the other hand,
have shown that frequency redistribution due to residual thermal
motion of atoms in the cold gas plays a significant role in
interpreting off resonance light diffusion in an ultracold atomic
sample.

Supported by the National Science Foundation (NSF-PHY-0355024), by
the Russian Foundation for Basic Research (RFBR-05-02-16172-a),
and by the North Atlantic Treaty Organization (PST-CLG-978468).


\begin{thebibliography}{99}
\bibitem{Metcalf} Harold J. Metcalf and Peter van der Straten,
\textit{Laser Cooling and Trapping} (Springer, New York, 1999).

\bibitem{LabeyrieOscillations} G. Labeyrie, Michaud, R. Kaiser,
submitted to Phys. Rev. Lett. (2005).

\bibitem{Holstein} T. Holstein, Phys. Rev. 72, 1212 (1947).

\bibitem{Fioretti} A. Fioretti, A.F. Molisch, J.H. Muller, P.
Verkerk, M. Allegrini, Opt. Comm. 149, 415 (1998).

\bibitem{LabeyrieTime1} G. Labeyrie, E. Vaujour, C.A. Muller, D.
Delande, C. Miniatura, D. Wilkowski, R. Kaiser, Phys. Rev. Lett.
91, 223904 (2003).

\bibitem{LabeyrieTime2} G. Labeyrie, R. Kaiser, and D. Delande,
Appl. Phys. B 81, 1001 (2005).

\bibitem{Balik1} S. Balik, R. G. Olave, C. I. Sukenik, and M. D.
Havey, Phys. Rev. A 72, 051402(R) (2005).

\bibitem{Sheng} P. Sheng, \textit{Introduction to Wave Scattering,
Localization, and Mesoscopic Phenomena} (Academic Press, San
Diego, 1995).

\bibitem{LagTig} A. Lagendijk, B.A. van Tiggelen, \textit{Resonant
Multiple Scattering of Light}, Phys. Rep. 270, 143 (1996).

\bibitem{Wiersma1} D.S. Wiersma, P. Bartolini, Ad Lagendijk,
R. Righini, Nature 390, 671 (1997).

\bibitem{Chabanov1} A.A. Chabanov, M. Stoytchev, A.Z. Genack, Nature
404, 850 (2000).

\bibitem{Mish} M.I. Mishchenko, Astrophys. J. 411, 351 (1993).

\bibitem{Labeyrie1} G. Labeyrie, F. De Tomasi, J-C Bernard, C.A. M\"{u}ller,
Ch. Miniatura, R. Kaiser, Phys. Rev. Lett. 83, 5266 (1999).

\bibitem{Labeyrie2} G. Labeyrie, C.A. Muller, D.S. Wiersma, Ch. Miniatura,
and R. Kaiser, J. Opt. B: Quantum Semiclass. Opt 2, 672 (2000).

\bibitem{Antilocalization}D.V. Kupriyanov, I.M. Sokolov, M.D.
Havey, Optics Comm. 243, 165 (2004); D.V. Kupriyanov, N.V.
Larionov, I.M. Sokolov, and M.D. Havey Optics and Spectroscopy 99
(3), 362 (2005).

\bibitem{HaveyCBS} D.V. Kupriyanov, I.M. Sokolov, P. Kulatunga,
C.I. Sukenik, M.D. Havey, Phys. Rev. A 67, 013814 (2003).

\bibitem{Ishimaru} J. Ishimaru and Yu. Kuga, J. Opt. Soc. Am. A1, 813 (1984).

\bibitem{Wolf} P.E. Wolf and G. Maret, Phys. Rev. Lett. 55, 2696 (1985).

\bibitem{Albeda} M.P. VanAlbada and A. Lagendijk, Phys. Rev. Lett. 55, 2692
(1985).

\bibitem{CommentAMO} Mark D. Havey and Dmitriy V. Kupriyanov,
Physica Scripta 72, C30,(2005).

\bibitem{LPRReview} D.V. Kupriyanov, I.M. Sokolov, C.I. Sukenik, and M.D.
Havey, Las. Phys. Lett. 3, 223 (2006).

\bibitem{Anderson} P.W. Anderson, Phys. Rev. 109, 1492 (1958).

\bibitem{Lukin1} M.D. Lukin, Rev. Mod. Phys. 75, 457 (2003).

\bibitem{Arimondo1} E. Arimondo and G. Orriols, Nuovo Cimento
Lett. 17, 333 (1976).

\bibitem{Gray} H.R. Gray, R. M. Whitley, and C.R. Sroud, Opt.
Lett. 3, 218 (1978).

\bibitem{Yoo} H.I. Yoo and J.H. Eberly, Phys. Rep. 118, 239
(1985).

\bibitem{Arimondo2} in Progress in Optics, ed. E. Wolf (Elsevier Science,
Amsterdam, 1996), Vol. XXXV, p. 257).

\bibitem{Harris} S.E. Harris, Phys. Today 50, 36 (1997).

\bibitem{Marangos} J.P. Marangos, J. Mod. Opt. 45, 471 (1998)

\bibitem{Wynands} R. Wynands and A. Nagel, Appl. Phys. B 68, 1 (1999)
.
\bibitem{Slowlight}L.V. Hau, S.E. Harris, Z. Dutton,
C.H. Behroozi, Nature 397, 594 (1999).

\bibitem{Braje} Danielle A. Braje, Vlatko Baic, Sunil Goda, G.Y.
Yin, and S.E. Harris, Phys. Rev. Lett. 93, 183601 (2004).

\bibitem{Liu2} Chien Liu, Zachary Dutton, Cyrus H. Behroozi, and
Lene Vestergaard Hau, Nature 409, 490 (2001).

\bibitem{Braje2} Danielle A. Braje, Vlatko Balic, G.Y. Yin, and
S.E. Harris, Phys. Rev. A 68, 041801 (2003).

\bibitem{Chaneliere} T. Chanelière, D. N. Matsukevich, S. D.
Jenkins, S.-Y. Lan, T. A. B. Kennedy, and A. Kuzmich, Nature 438,
833 (2005).

\bibitem{Eisaman} M. D. Eisaman, A. André, F. Massou, M.
Fleischhauer , A. S. Zibrov, and M. D. Lukin, Nature 438, 837
(2005).

\bibitem{Yan} Min Yan, Edward G. Rickey, and Yifu Zhu, J. Opt.
Soc. Am. B 18, 1057 (2001).

\bibitem{note} Note that in warm atomic vapors, thermal atomic
motion washes out the interferences.  A.B. Matsko, I. Novikova,
M.O. Scully, and G.R. Welch, Phys. Rev. Lett. 87, 133601 (2001).

\bibitem{Sokolov1} D.V. Datsyuk, I.M. Sokolov, JETP,
\textbf{129} 1-14 (2006).

\end{thebibliography}
\end{document}